\documentclass[aip,apl,preprint,floatfix,amsfonts,amsmath,amssymb]{revtex4-1}

\usepackage{graphicx,dcolumn,bm}

\begin{document}

\title{Inducing out-of-plane precession of magnetization for microwave assisted magnetic recording using an oscillating polarizer in spin torque oscillator}

\author{W. Zhou}
\email{ZHOU.Weinan@nims.go.jp}
\affiliation{Research Center for Magnetic and Spintronic Materials, National Institute for Materials Science (NIMS), Tsukuba 305-0047, Japan}

\author{H. Sepehri-Amin}
\affiliation{Research Center for Magnetic and Spintronic Materials, National Institute for Materials Science (NIMS), Tsukuba 305-0047, Japan}

\author{T. Taniguchi}
\affiliation{National Institute of Advanced Industrial Science and Technology (AIST), Spintronics Research Center, Tsukuba 305-8568, Japan}

\author{S. Tamaru}
\affiliation{National Institute of Advanced Industrial Science and Technology (AIST), Spintronics Research Center, Tsukuba 305-8568, Japan}

\author{Y. Sakuraba}
\email{SAKURABA.Yuya@nims.go.jp}
\affiliation{Research Center for Magnetic and Spintronic Materials, National Institute for Materials Science (NIMS), Tsukuba 305-0047, Japan}

\author{S. Kasai}
\affiliation{Research Center for Magnetic and Spintronic Materials, National Institute for Materials Science (NIMS), Tsukuba 305-0047, Japan}

\author{H. Kubota}
\affiliation{National Institute of Advanced Industrial Science and Technology (AIST), Spintronics Research Center, Tsukuba 305-8568, Japan}

\author{K. Hono}
\affiliation{Research Center for Magnetic and Spintronic Materials, National Institute for Materials Science (NIMS), Tsukuba 305-0047, Japan}

\date{\today}

\begin{abstract}
We investigated the dynamics of a novel design of spin torque oscillator (STO) for microwave assisted magnetic recording.
Using Ni$_{80}$Fe$_{20}$ (NiFe) as the polarizer and Fe$_{67}$Co$_{33}$ (FeCo) as the field generating layer, we experimentally observed the magnetization reversal of NiFe, followed by multiple signals in the power spectra as the bias voltage increased.
The signals reflected the out-of-plane precession (OPP) mode oscillation of both FeCo and NiFe, as well as the magnetoresistance effect of the STO device, which had the frequency equal to the difference between the oscillation frequency of NiFe and FeCo.
Such dynamics were reproduced by micromagnetic simulation.
In addition to the merit of realizing the OPP mode oscillation with a simple and thin structure suitable for a narrow gap recording head, the experimental results using this design suggested that a large cone angle of $\sim$ 70$^{\circ}$ for the OPP mode oscillation of FeCo was achieved, which was estimated based on the macrospin model.
\end{abstract}


\maketitle

Energy assisted recording technologies, \textit{e.g.}, microwave assisted magnetic recording (MAMR), have become indispensable to maintain the continuous growth of recording density of hard disk drives.\cite{MAMR1,HDD}
In order to fulfill the requirements of signal-to-noise ratio and thermal stability simultaneously, materials with increasing magnetocrystalline anisotropy ($K_{\text{u}}$) are being exploited as recording media.
MAMR grants writability to high $K_{\text{u}}$ media by applying an additional ac magnetic field  ($h_{\text{ac}}$) to induce the precessional motion of magnetic moments, which results in magnetization switching under a much smaller magnetic field ($H$) than the coercivity.\cite{mas1}
One technical challenge for MAMR is to generate high frequency, large amplitude $h_{\text{ac}}$ within a nanosized area.
It has been shown that for the media with the effective anisotropy field of 2 T, $h_{\text{ac}}$ with the frequency of 18 GHz and the amplitude of 0.1 T is necessary for a sufficient switching field reduction.\cite{mas1}
However, higher values are required as we pursuing for higher recording density.
The $h_{\text{ac}}$ generation is expected to be realized with a spin torque oscillator (STO).\cite{OPP1,STOrev1,STOrev2,STOrev3,mas2}
The STO device is placed in the narrow gap between main pole and trailing shield of the recording head, and a perpendicular $H$ of $\sim$ 1 T is applied to the device during recording.
As the current passes through, the spin-polarized electrons apply spin-transfer torque (STT)\cite{STT1, STT2} to one soft magnetic layer to cancel the damping torque, which makes the magnetization undergo the out-of-plane precession (OPP) mode oscillation\cite{OPP1, OPP2} for $h_{\text{ac}}$ generation.
This layer is called the field generating layer (FGL).
Previously, it is proposed to use a perpendicularly magnetized layer as the polarizer for stable oscillation.\cite{pSTO1,pSTO2,pSTO3}
The STO devices using a 3-nm-thick Co$_2$Fe(Ga$_{0.5}$Ge$_{0.5}$) layer that is exchange-coupled with a 10-nm-thick \textit{L}1$_0$-FePt as the polarizer were experimentally studied, and the combination of the materials with high spin polarization and high $K_{\text{u}}$ showed oscillation performance close to the requirement of practical MAMR application.\cite{pSTO4,pSTO5}
However, its thick structure requires a wide gap between main pole and trailing shield, which leads to broadening of the field distribution from main pole and results in recording transition noise.\cite{mainpole}

\begin{figure}
\includegraphics{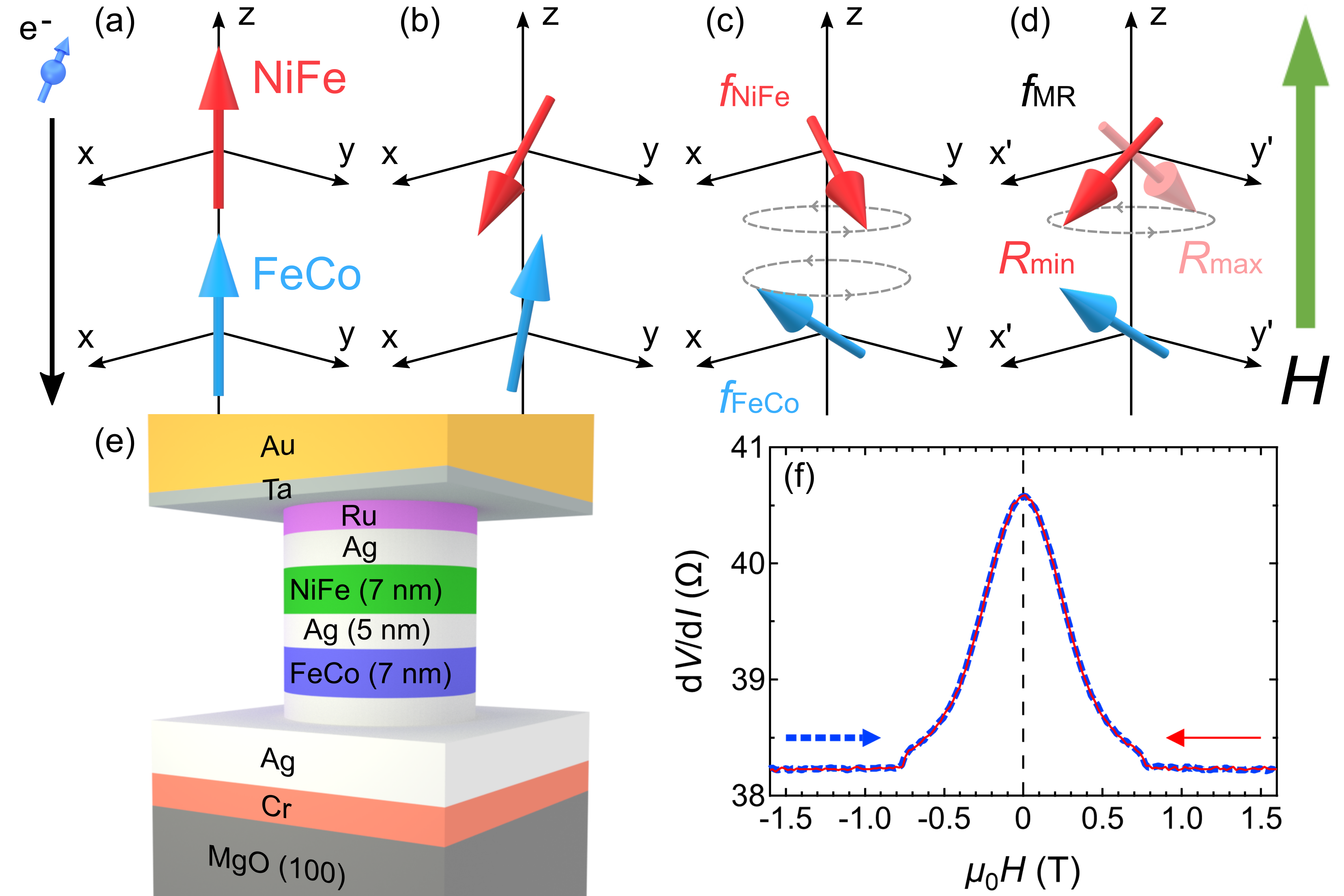}
\caption{\label{Fig1} (a) Schematic illustration of the magnetization of both NiFe and FeCo aligned along $H$. NiFe was used as the polarizer and FeCo as the FGL. (b) NiFe was reversed due to STT. (c) Both NiFe and FeCo were in the OPP mode oscillation. (d) If the coordinate rotates with FeCo along the z-axis, then in this coordinate (x' and y'), FeCo stays still while NiFe oscillates with frequency equal to $f_{\text{NiFe}} - f_{\text{FeCo}}$, which is also the frequency of the change of resistance due to MR effect ($f_{\text{MR}}$). (e) Schematic illustration of the circular pillar of the STO device. (f) MR curve of the STO device under perpendicular $H$ and $U$ of $- 1$ mV. The arrows indicates the $H$ sweeping direction of the corresponding curves.}
\end{figure}

Recently, Zhu \textit{et al.} proposed a novel design of STO, where only a soft magnetic thin layer is exploited as the polarizer.\cite{iSTO1,iSTO2}
Under perpendicular $H$, the magnetization of both the FGL and polarizer will align along $H$, as schematically illustrated in Fig.~\ref{Fig1}(a).
As the electrons flow from top to bottom, the antiparallel configuration of magnetization is favored, and STT will first reverse the polarizer (Fig.~\ref{Fig1}(b)).
Then,  the electron will be spin-polarized to the direction of polarizer (opposite to $H$) and apply STT to FGL to induce the OPP mode oscillation.
This operation mechanism no longer requires a perpendicularly magnetized polarizer or layer to pin the polarizer, makes it possible to generate $h_{\text{ac}}$ with a thin and simple STO device.
The dynamics of STO devices consisting of only soft magnetic layers have been studied under large perpendicular $H$ and high current density.\cite{OPP1,STOref1,STOref2,STOref3}
However, the OPP mode oscillation has not been clearly demonstrated and discussed from the viewpoint of MAMR application.

In this letter, we report on the experimental demonstration of the OPP mode oscillation using the aforementioned mechanism.
Our results showed that the polarizer and FGL were both in the OPP mode oscillation with different frequency (Fig.~\ref{Fig1}(c)).
And the change of resistance due to the magnetoresistance (MR) effect had the frequency equal to the difference between those of the polarizer and FGL (Fig.~\ref{Fig1}(d)).
Such dynamics were reproduced by micromagnetic simulation.
We also show the estimated cone angle ($\theta$) of the OPP mode oscillation based on the macrospin model.

We used Fe$_{67}$Co$_{33}$ (FeCo) as the FGL due to its large saturation magnetization ($M_s$), and Ni$_{80}$Fe$_{20}$ (NiFe) as the spin polarizer.
The STO devices were microfabricated from the blanket thin film with the stacking structure of MgO (100) subs.~// Cr (10) / Ag (100) / FeCo (7) / Ag (5) / NiFe (7) / Ag (5) / Ru (8) (thickness in nanometers).
Schematic illustration of the circular pillar of STO is shown in Fig.~\ref{Fig1}(e).
Detailed descriptions of the fabrication process of STO can be found in the Supplemental
Material.
Because the small pillars were covered by thick electrodes and could not be clearly observed in scanning electron microscope (SEM), we estimated the pillar diameter using the SEM measured diameter of large size pillars ($D \sim$ 140 and 350 nm) on the same sample, and the change of resistance (${\Delta}R$) obtained from the MR curves, based on the linear relationship between ${\Delta}R$ and the reciprocal of the area of the pillar (${\Delta}R \propto 1/A$).
The experimental results reported here were measured from a device with the diameter of $\sim$ 28 nm.
During the measurement, the sample was mounted on a sample fixture having a 2-axis rotary stage and equipped with a custom high frequency probe, which was inserted into an electromagnet.
This setup allowed us to apply external $H$ along arbitrary directions.\cite{setup}
In the power spectral density (PSD) measurement, a bias DC voltage ($U$) was applied to the STO device through a bias-tee.
The generated signal was amplified by a low noise amplifier, and captured by a commercial spectrum analyzer.
We did not subtract the amplifier gain from the results of PSD.
A lock-in amplifier in addition to the DC voltage source was connected to the DC port during the measurement of $R$ and d$V$/d$I$.
The positive voltage and current density was defined as electrons flowing from the top NiFe layer to the bottom FeCo layer.
All the measurements were carried out at room temperature.

\begin{figure}
\includegraphics{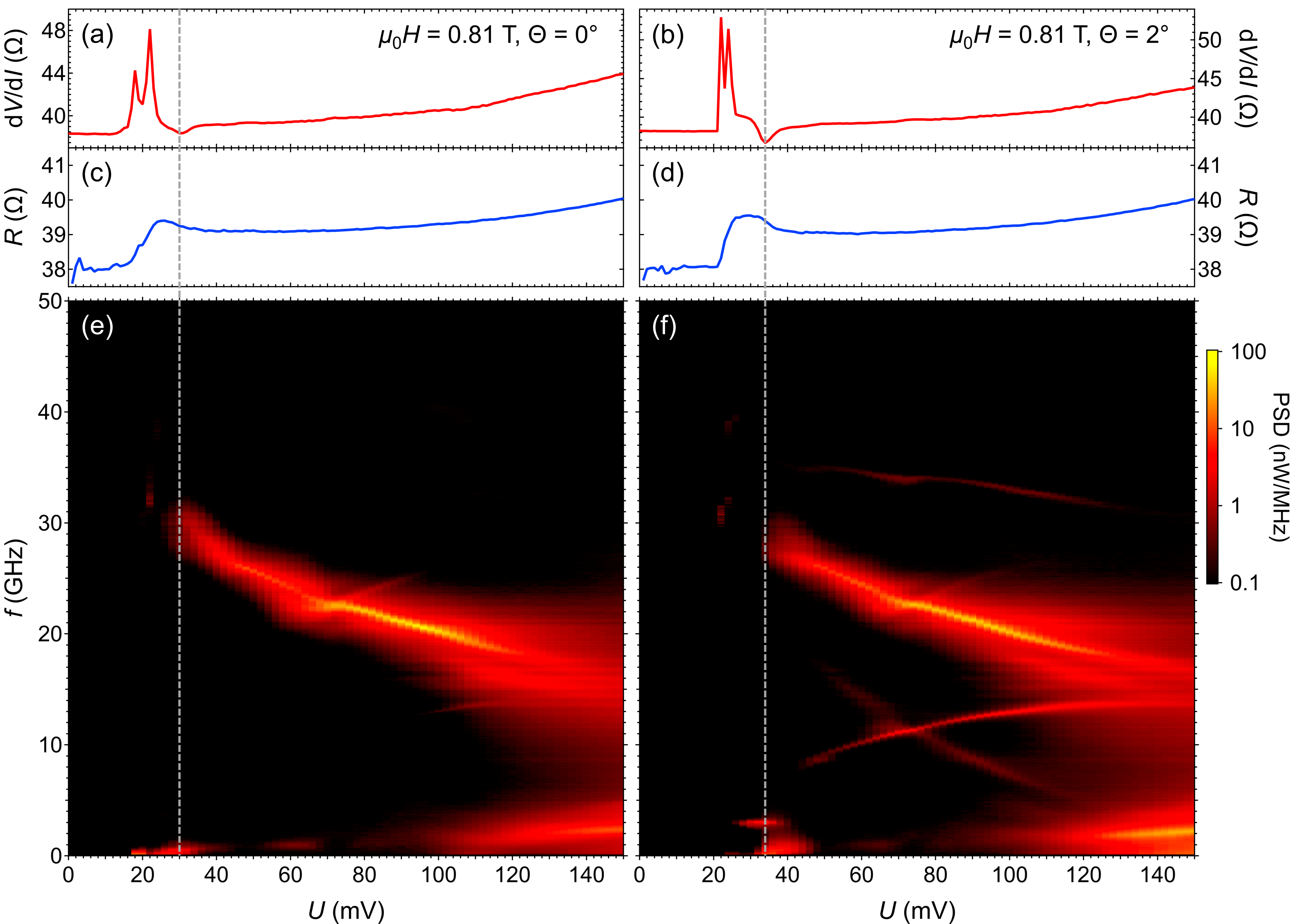}
\caption{\label{Fig2} (a) d$V$/d$I$ of the STO device as a function of $U$ under perpendicular $\mu_{0}H$ = 0.81 T along $\Theta = 0^{\circ}$ and (b) $\Theta = 2^{\circ}$ direction. (c) $R$ as a function of $U$ under perpendicular $\mu_{0}H$ = 0.81 T along $\Theta = 0^{\circ}$ and (d) $\Theta = 2^{\circ}$ direction. (e) The mappings of PSD under perpendicular $\mu_{0}H$ = 0.81 T along $\Theta = 0^{\circ}$ and (f) $\Theta = 2^{\circ}$ direction.}
\end{figure}

Figure~\ref{Fig1}(f) shows the MR curve of the STO device under perpendicular $H$ and a low $U$ of $- 1$ mV.
At zero $H$, the NiFe and FeCo layers have their magnetization in-plane with an antiparallel configuration due to the dipole-dipole interaction, resulting in a high $R$ state.
As $H$ increases along the perpendicular direction, the magnetization of both layers aligns towards $H$, and $R$ gradually decreases.
The MR ratio of this device is $\sim$ 6.2 \%.
In addition, $R$ reaches the minimum at $\mu_{0}H <$ 1 T,  which is much smaller than the $M_\text{s}$ of FeCo, indicating a large reduction of the demagnetization factor due to the small lateral size of the pillar.
Figures~\ref{Fig2}(a), (c) and (e) show the d$V$/d$I$ and R as a function of $U$, and the mappings of the PSD of the STO device under perpendicular $\mu_{0}H$ = 0.81 T (the angle between $H$ and z-axis, \textit{i.e.}, $\Theta = 0^{\circ}$).
For the measurement, $H$ was held still while $U$ was increased from 0 to 150 mV.
Here, the magnetization of the NiFe layer is reversed at $U \sim$ 22 mV, which is indicated by the peaks in the d$V$/d$I$ curve as well as the increase in $R$.\cite{STOref1,STOref2,STOref3,STOref4}
After that, there is a small dip in the d$V$/d$I$ curve at $U$ = 30 mV marked by the gray dash line, which is the threshold $U$ for the appearance of the strong microwave signal.
The dip in the d$V$/d$I$ curve corresponds to the decrease in $R$, which was also observed in previous studies,\cite{STOref2,STOref3} and is attributable to the dynamics excitation of the FeCo layer.
As $U$ increases, the frequency of the strong microwave signal decreases (red-shift).
The same measurement was also carried out with the same value of $H$ tilted to $\Theta = 2^{\circ}$, and the results are shown in Figs.~\ref{Fig2}(b), (d) and (f).
In the d$V$/d$I$ curve, the peaks and dip shift towards higher values of $U$ $\sim$ 24 mV and 34 mV, respectively.
For the mapping of PSD, a similar strong signal was observed after the dip in the d$V$/d$I$ curve, together with other weak signals.
Here, we emphasize the appearance of the weak signal with frequency higher than the strong signal, and the one with frequency lower than the strong signal having the frequency increased as $U$ increased (blue-shift).

\begin{figure}
\includegraphics{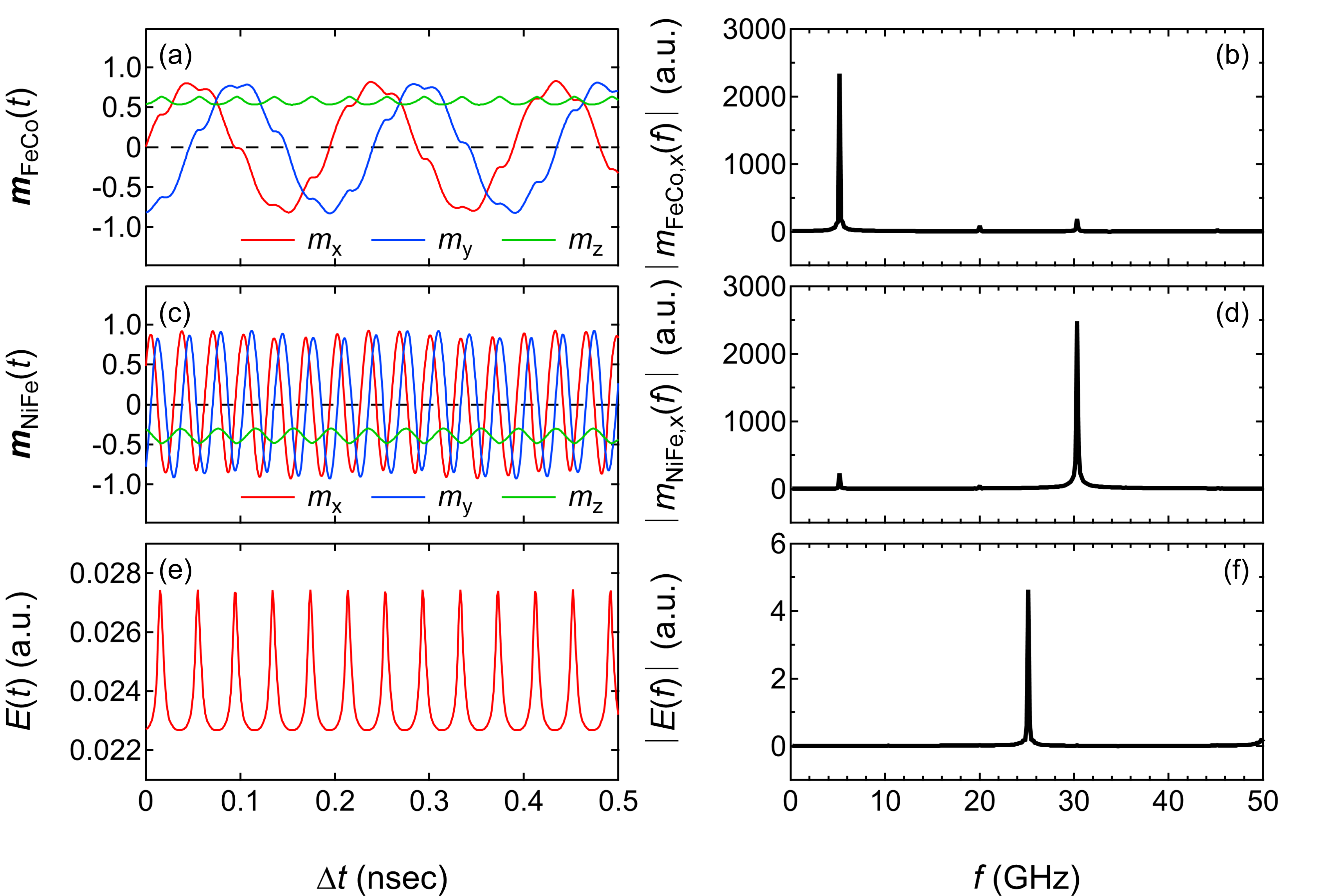}
\caption{\label{Fig3} (a) Time evolution of $\bm{m}_\text{FeCo}$ in a stable oscillation state under perpendicular $\mu_{0}H$ = 0.81 T and $J$ = 3.2 $\times$ 10$^{8}$ A/cm$^{2}$ obtained from micromagnetic simulation. (b) DFT magnitude of the x component of $\bm{m}_\text{FeCo}$. (c) Time evolution of $\bm{m}_\text{NiFe}$. (d) DFT magnitude of the x component of $\bm{m}_\text{NiFe}$. (e) Time evolution of the corresponding $E$. (f) DFT magnitude of $E$.}
\end{figure}

To better understand the dynamics, micromagnetic simulation was carried out using the software $magnum.fe$,\cite{sim1} which has the capability to calculate the coupled dynamics of magnetization and the spin accumulation simultaneously by solving the Landau-Lifshitz-Gilbert (LLG) equation and the time dependent 3D spin diffusion equation.
A 28-nm-diameter circular pillar consisting of a 7-nm-thick NiFe layer and a 7-nm-thick FeCo layer separated by a 5-nm-thick non-magnetic layer, was employed as the model for simulation.
The $\mu_{0}M_s$, exchange stiffness ($A$) and spin polarization ($\beta$) of NiFe were set as 1.0 T, 13 pJ/m,\cite{paraNiFe} and 0.4, respectively; while $\mu_{0}M_s$ = 2.3 T, $A$ = 30 pJ/m,\cite{paraFeCo} and $\beta$ = 0.5 were used for FeCo.
The damping constant ($\alpha$) of 0.01 was used for both NiFe and FeCo.
Figures~\ref{Fig3}(a) and (c) show the time evolution of the averaged magnetization vector ($\bm{m}$) of the FeCo and NiFe layers, respectively, in a stable oscillation state under perpendicular $\mu_{0}H$ = 0.81 T and current density ($J$) of 3.2 $\times$ 10$^{8}$ A/cm$^{2}$ (this $J$ approximately corresponds to $U$ = 80 mV in experiment), while the electrical potential ($E$) between the top and bottom of the circular pillar is shown in Fig.~\ref{Fig3}(e).
The time evolution of the x and y components of $\bm{m}_\text{FeCo}$ and $\bm{m}_\text{NiFe}$ indicates that both the FeCo and NiFe layers are in the OPP mode oscillation, and oscillate in the same direction.
The z component is positive for $\bm{m}_\text{FeCo}$ while negative for $\bm{m}_\text{NiFe}$, indicating that the magnetization of the NiFe layer is reversed.
Using the discrete Fourier transform (DFT), the corresponding spectra in frequency domain were calculated and shown in Figs.~\ref{Fig3}(b), (d) and (f).
For the FeCo layer, the largest magnitude appears at 5.17 GHz, which is the frequency of the OPP mode oscillation of FeCo ($f_\text{FeCo}$); for the NiFe layer, $f_\text{NiFe}$ = 30.33 GHz.
On the other hand, for the spectrum of $E$, which corresponds to the experimentally measured PSD,  the largest magnitude appears at $f_\text{MR}$ = 25.16 GHz.
This value is different from either $f_\text{FeCo}$ or $f_\text{NiFe}$, however, equal to the difference between $f_\text{NiFe}$ and $f_\text{FeCo}$, \textit{i.e.}, $f_\text{MR} = f_\text{NiFe} - f_\text{FeCo}$.

\begin{figure}
\includegraphics{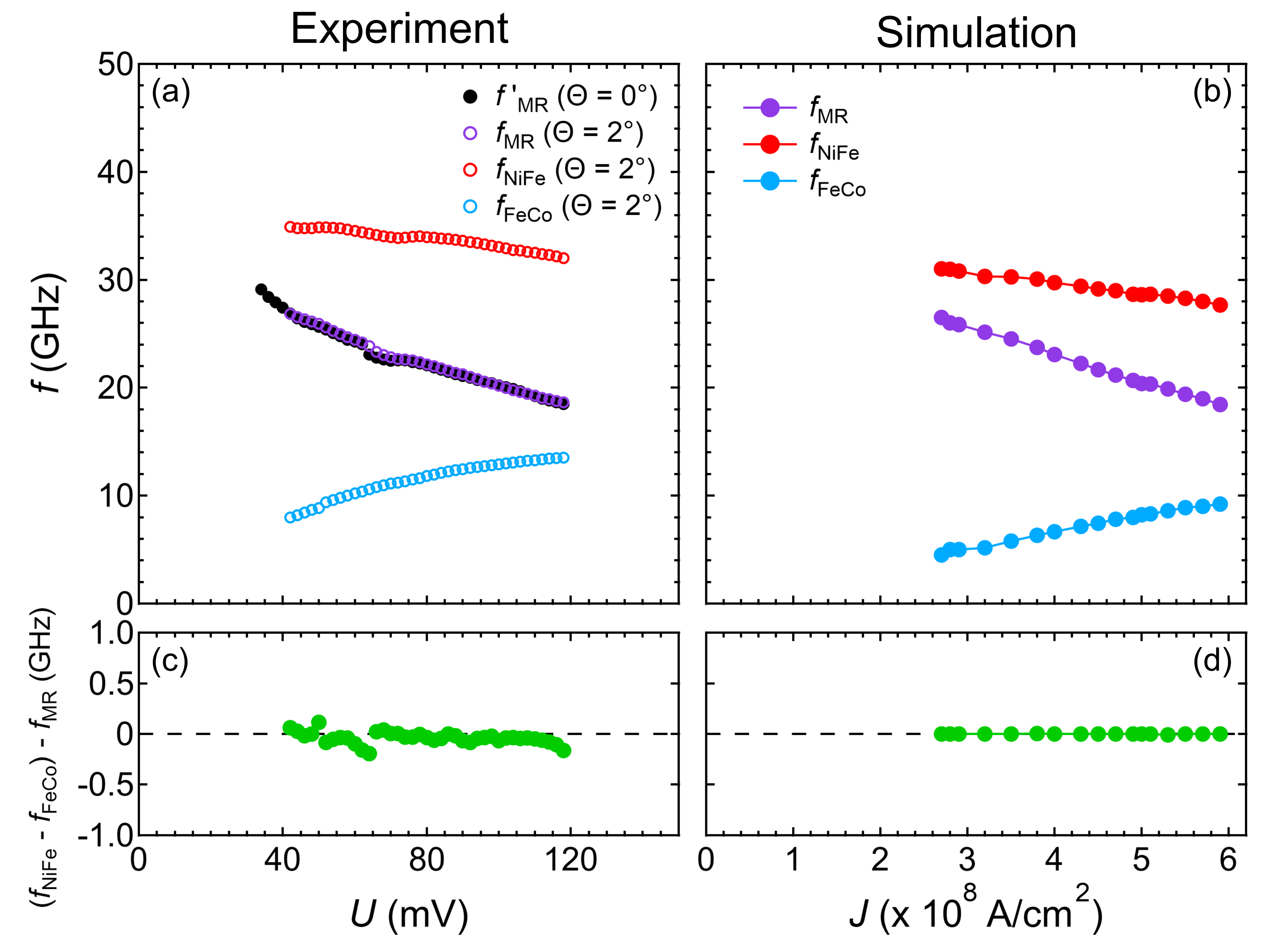}
\caption{\label{Fig4} (a) Frequency of some of the microwave signals in PSD from Figs.~\ref{Fig2}(e) and (f) as a function of $U$. (b) Frequency of the largest magnitude in the spectra of x components of $\bm{m}_\text{NiFe}$ and $\bm{m}_\text{FeCo}$ as well as $E$ as a function of $J$. (c) $(f_\text{NiFe} - f_\text{FeCo}) - f_\text{MR}$ from experiment as a function of $U$ and (d) from simulation as a function of $J$. $U$ = 80 mV in experiment corresponds to $J =$ 3.3 $\times$ 10$^{8}$ A/cm$^{2}$.}
\end{figure}

This relationship is also observed experimentally.
The frequency of some of the microwave signals in PSD as a function of $U$ from Figs.~\ref{Fig2}(e) and (f) was extracted and is shown in Fig.~\ref{Fig4}(a).
For the strong red-shift signals, the ones of $\Theta = 2^{\circ}$ (purple hollowed circles) overlap with the ones of $\Theta = 0^{\circ}$ (black circles) with little deviation, indicating there is no fundamental change in the oscillation dynamics caused by the tilting of $H$.
Furthermore, the previously emphasized signals plotted as the red and blue hollowed circles, have the differences of frequency equal to that of the strong red-shift signals, as exhibited in Fig.~\ref{Fig4}(c).
The corresponding results from simulation under $\mu_{0}H$ = 0.81 T is shown in Figs.~\ref{Fig4}(b) and (d).
Here, we plotted the frequency of the largest magnitude in the spectra in Figs.~\ref{Fig3}(b), (d) and (f) as a function of $J$.
The comparison between experiment and simulation indicates that the strong red-shift signal in experiment is due to the MR effect of the STO device as schematically illustrated in Fig.~\ref{Fig1}(d).
The signals represented by the red and blue hollowed circles reflect the OPP mode oscillation of the NiFe and FeCo layers, respectively.
The appearance of the signals of$f_\text{NiFe}$ and $f_\text{FeCo}$ might be caused by the distortion of the OPP mode oscillation trajectory due to the tilting of $H$, which led to change of $R$ in every period of oscillation.
The red-shift of $f_\text{MR}$ and $f_\text{NiFe}$ and the blue-shift of $f_\text{FeCo}$ are qualitatively reproduced in simulation.

\begin{figure}
\includegraphics{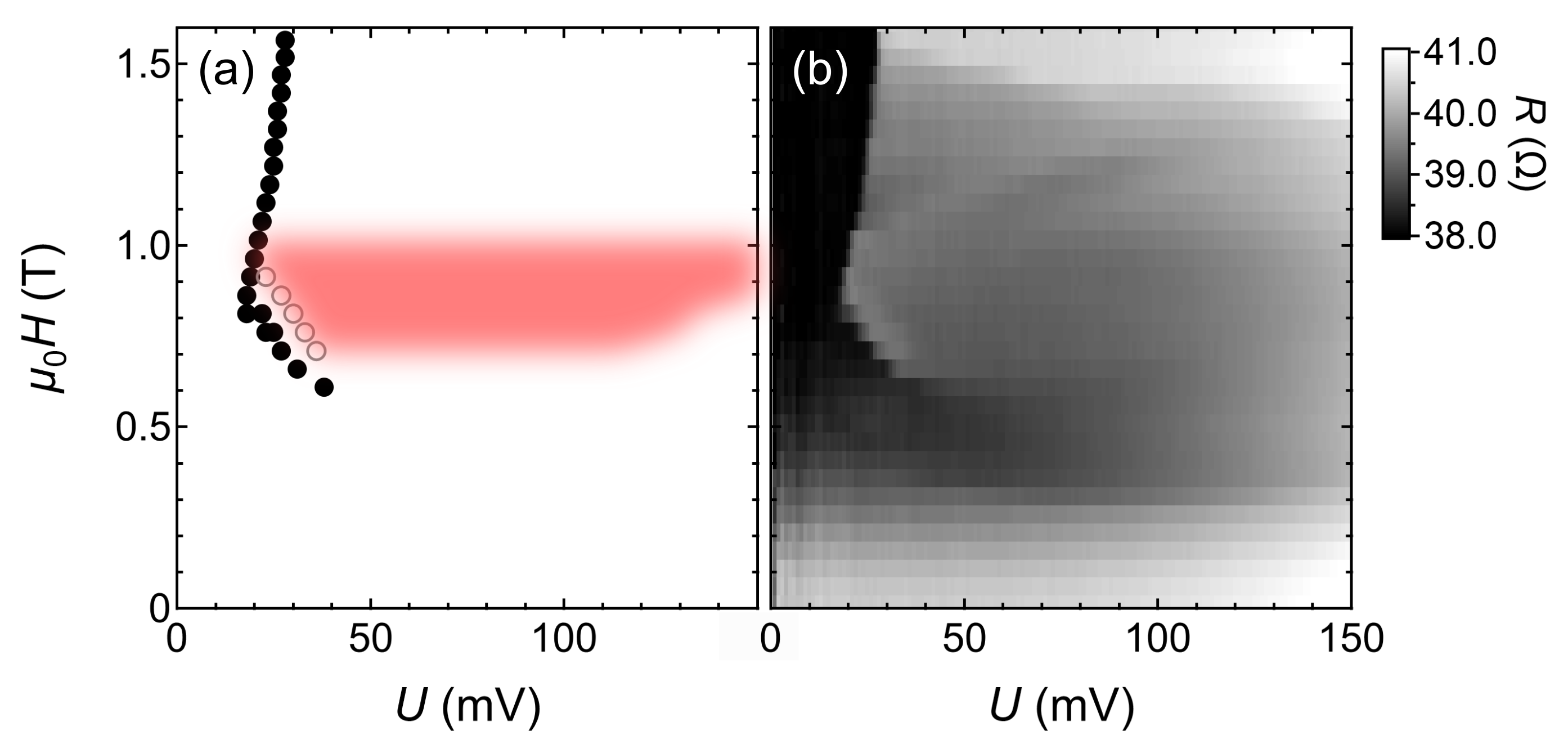}
\caption{\label{Fig5} (a) Peaks (black circles) and dips (gray hollowed circles) in the d$V$/d$I$ curves mapped on the $U$ - $H$ plane. The red area marks the condition where the signals due to the OPP mode oscillation of both layer was observed from PSD. (b) $R$ of the device mapped on the $U$ - $H$ plane. $H$ in (a) and (b) was along the perpendicular direction ($\Theta = 0^{\circ}$).}
\end{figure}

Some of the peaks and dips from the d$V$/d$I$ curves and $R$ of the STO device were mapped on the $U$ - $H$ plane as shown in Figs.~\ref{Fig5}(a) and (b), respectively.
At zero $H$, the device shows high $R$ state.
As $H$ increases, $R$ decreases to the minimum value under low $U$, however, suddenly increases as $U$ increases to $\sim$ 20 mV.
This behavior corresponds to the peaks in the d$V$/d$I$ curves (black circles in Fig.~\ref{Fig5}(a)), which is caused by the reversal of the NiFe layer.
After the reversal of NiFe, under high $\mu_{0}H$ $\sim$ 1.5 T, the device shows high value of $R$ close to the ones under zero $H$, indicating that NiFe and FeCo are antiparallel, however, along the z-axis.\cite{STOref4}
In the region between the high and low $H$, the device shows intermediate $R$.
And within this region, the area colored red in Fig.~\ref{Fig5}(a) is the condition where signal due to the OPP mode oscillation of both layer was observed from PSD.
The threshold $U$ on the left side of the red area coincides with the dips in the d$V$/d$I$ curves, as marked by the gray hollowed circles in Fig.~\ref{Fig5}(a).

\begin{figure}
\includegraphics{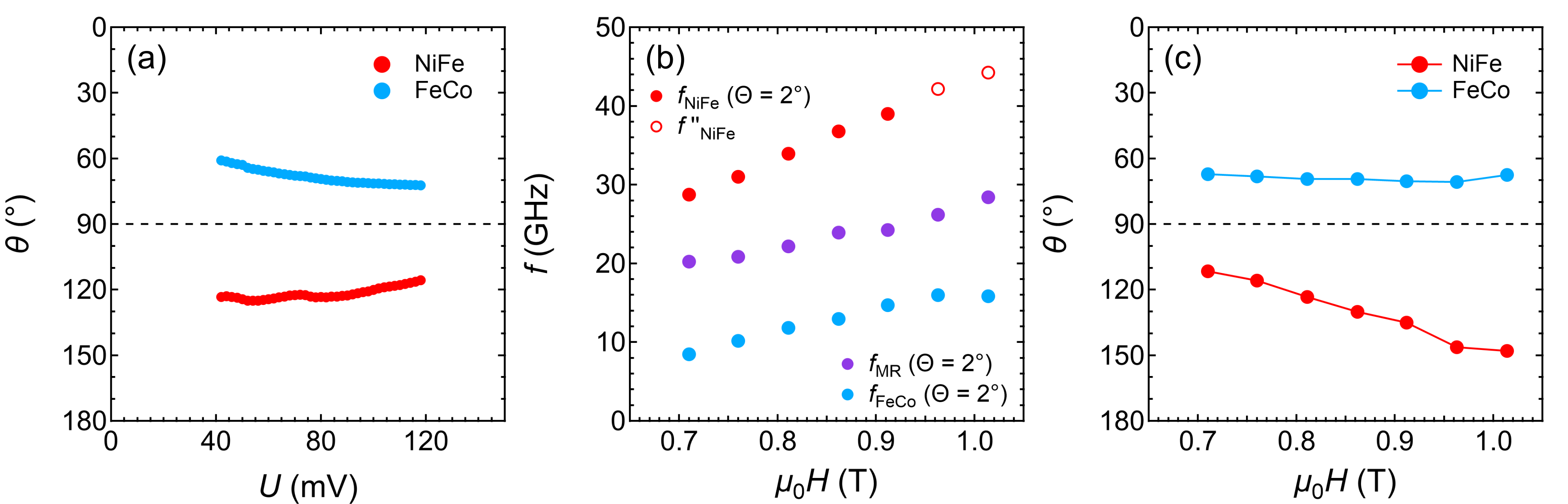}
\caption{\label{Fig6} (a) The estimated $\theta$ of NiFe and FeCo using the results from Fig.~\ref{Fig4}(a). (b) $f_\text{NiFe}$, $f_\text{FeCo}$, and $f_\text{MR}$ as a function of $H$ under $U$ = 80 mV. $f''_\text{  NiFe}$ is calculated using the relationship of $f_\text{MR} = f_\text{NiFe} - f_\text{FeCo}$. (c)  The estimated $\theta$ of NiFe and FeCo using the results from (b).}
\end{figure}

For practical application, a large $\theta$ for the OPP mode oscillation is important, since it determines the amplitude of generated $h_{\text{ac}}$.
We estimated $\theta$ from the positive z-axis of both NiFe and FeCo ($\theta_\text{NiFe}$ and $\theta_\text{FeCo}$) based on $f_\text{NiFe}$ and $f_\text{FeCo}$ obtained from experiment.
We employed the macrospin model, and assumed that the frequency of the OPP mode oscillation is proportional to the effective field of the layer, which is the sum of the external $H$, the demagnetizing field, and the dipole field generated from the other layer.\cite{Hd} (see Supplemental Material for the detailed description of the estimation)
Figure~\ref{Fig6}(a) shows the estimated $\theta$ using the results from Fig.~\ref{Fig4}(a).
A large $\theta_\text{FeCo} \sim$ 60$^{\circ}$ appears at $U \sim$ 40 mV, and gradually increases to $\sim$ 70$^{\circ}$ as $U$ increases.
This is attributable to the increase of STT with increasing $J$.
On the other hand, $\theta_\text{NiFe} \sim$ 120$^{\circ}$, and slightly decreases as $U$ increases.
The field-dependence of $\theta$ was also investigated.
Figure~\ref{Fig6}(b) shows the $f_\text{NiFe}$, $f_\text{FeCo}$, and $f_\text{MR}$ as a function of $H$ obtained from the PSD under $U$ = 80 mV, which corresponds to $J =$ 3.3 $\times$ 10$^{8}$ A/cm$^{2}$.
At high $H \sim$ 1 T, the signal of $f_\text{NiFe}$ was so weak that cannot be distinguished in PSD, and we used the values calculated from the relationship of $f_\text{MR} = f_\text{NiFe} - f_\text{FeCo}$, as indicated by the red hollowed circles.
$f_\text{NiFe}$, $f_\text{MR}$, and $f_\text{FeCo}$ increase as $H$ increases, and $f_\text{FeCo}$ shows a maximum value of $\sim$ 16 GHz.
The estimation of $\theta$ using these results is shown in Fig.~\ref{Fig6}(c).
Here, $\theta_\text{FeCo}$ exhibits the values of $\sim$ 70$^{\circ}$ with small changes due to $H$, while $\theta_\text{NiFe}$ increases as $H$ increased.
It is worthwhile to mention that for the mechanism of STO studied here, because the polarizer has its magnetization reversed opposite to $H$ (Fig.~\ref{Fig1}(c)), its demagnetizing field has positive z component while negative z component for FGL.
This leads to usually higher effective field for polarizer, thus higher frequency of the OPP mode oscillation than that of FGL.

In conclusion, we investigated the dynamics of the novel design of STO for MAMR, where only a soft magnetic thin layer is exploited as the polarizer.
Using NiFe as the polarizer and FeCo as the FGL, our results from experiment and simulation clearly show the OPP mode oscillation for both layers with different frequency as $f_\text{NiFe}$ and $f_\text{FeCo}$, respectively.
Such dynamics also generated the microwave signal with $f_\text{MR} = f_\text{NiFe} - f_\text{FeCo}$.
Based on the macrospin model, $\theta$ of the OPP mode oscillation was estimated, and the results suggested a large $\theta$ of $\sim$ 70$^{\circ}$ for the FeCo layer at high $f_\text{FeCo} \sim$ 16 GHz. 

See supplemental Material for the detailed description of the fabrication process of the STO devices, and the estimation of $\theta$.

This work was supported by the Advanced Storage Research Consortium (ASRC), Japan, Grant-in-Aids for Scientific Research (S) (Grant No. 17H06152) and Grant-in-Aids for Young Scientific Research (B) (Grant No. 17K14802) from the Japan Society for the Promotion of Science (JSPS).
The authors thank H. Suto, S. Tsunegi, T. M. Nakatani and R. Iguchi for valuable discussions, and N. Kojima for technical support.

\begin{center}
\textbf{\large Supplemental Material}
\end{center}

\begin{flushleft}
\textbf{1. Fabrication process of the STO devices}
\end{flushleft}

In the experiment, Fe$_{67}$Co$_{33}$ (FeCo) was exploited as the field generating layer (FGL) due to its large saturation magnetization ($M_s$), while Ni$_{80}$Fe$_{20}$ (NiFe) was used as the polarizer.
The blanket thin film with the stacking structure of MgO (100) subs.~// Cr (10) / Ag (100) / FeCo (7) / Ag (5) / NiFe (7) / Ag (5) / Ru (8) (thickness in nanometers) was prepared.
The Cr / Ag buffer layer was annealed at 300$^{\circ}$C for 30 min to increase the surface flatness, and the rest of the layers were deposited at room temperature.
For the fabrication process, circular pillars were first microfabricated using electron beam lithography and Ar ion milling.
The smallest patterned resist pillars had the diameter of $\sim$ 80 nm.
In order to make the STO devices with the diameter less than 40 nm required by practical application, during the milling process, the resist pillars were first milled at the angle of 75$^{\circ}$ from the normal direction of the sample to reduce the resist pillar size, before functioning as mask to form the circular pillars.
The milling process stopped within the 100-nm-thick Ag layer.
Then, SiO$_{2}$ was deposited to fill the space around the pillars for insulation, followed by the formation of the Ta (2) / Au (150) top electrodes.

\begin{flushleft}
\textbf{2. Estimation of the cone angle ($\theta$)}
\end{flushleft}

Since the size of the magnetic layers is small, and we did not observe stable oscillation of multi-domain state within the values of current density ($J$) in our micromagnetic simulation ($J <$ 6.1 $\times$ 10$^{8}$ A/cm$^{2}$), it is reasonable to employ the macrospin model and represent the magnetization of each layer with a single vector.
Under the stable OPP mode oscillation state as shown in Fig.~\ref{Fig1}(c), the oscillation frequency should be proportional to the effective field ($H_\text{eff}$) of the layer as $f = (\gamma / 2\pi) H_\text{eff}$, where $\gamma$ is the gyromagnetic ratio.
We assume $H_\text{eff}$ is the sum of the external field ($H_\text{ext}$), the demagnetizing field ($\bm{H}_\text{demag}$), and the dipole field ($H_\text{d}$) generated from the other layer.
$\bm{H}_\text{demag}$ is usually expressed as ($-M_{s}N_{x}m_{x}$, $-M_{s}N_{y}m_{y}$, $-M_{s}N_{z}m_{z}$), where $N_{x}$, $N_{y}$, and $N_{z}$ are the x, y, and z component of the demagnetization factor having $N_{x} + N_{y} + N_{z} = 1$.
However, for a circular pillar, $N_{x} = N_{y}$, so we can use $\bm{H}_\text{demag} + M_{s}N_{x}\bm{m} = (0, 0, -M_{s}(3N_{z} - 1)m_{z} / 2)$ instead, where only the z component remains.
And this correction will not affect the oscillation state, since in Landau-Lifshitz-Gilbert (LLG) equation, $\bm{m} \times \bm{H}_\text{demag} = \bm{m} \times (\bm{H}_\text{demag} + M_{s}N_{x}\bm{m})$.
For $H_\text{d}$, because FeCo and NiFe oscillated in different frequency, we only consider the z component of the field at the center of one layer generated by the magnetization of the other layer, and it is calculated using Eq.(14) in Ref.~\onlinecite{Hd}.
Based on these assumptions, we derived the expression of oscillation frequency as
\begin{equation}
f_\text{NiFe}=\frac{\gamma}{2\pi}(H_\text{ext}-\frac{3N_\text{z}-1}{2}M_\text{s,NiFe}\cos\theta_\text{NiFe}+H_\text{d,NiFe}\cos\theta_\text{FeCo}), \tag{1a}
\end{equation}
\begin{equation}
f_\text{FeCo}=\frac{\gamma}{2\pi}(H_\text{ext}-\frac{3N_\text{z}-1}{2}M_\text{s,FeCo}\cos\theta_\text{FeCo}+H_\text{d,FeCo}\cos\theta_\text{NiFe}), \tag{1b}
\end{equation}
where the average $\theta_\text{NiFe}$ and $\theta_\text{FeCo}$ of the OPP mode oscillation could be calculated.

\bibliography{Text_ref}

\end{document}